# Smart Grid Management using Blockchain: Future Scenarios and Challenges

Tudor Cioara
*Computer Science Department*
*Faculty of Automation and Computer Science*
*Technical University of Cluj-Napoca*
Cluj-Napoca, Romania
tudor.cioara@cs.utcluj.ro

Claudia Pop (Antal)
*Computer Science Department*
*Faculty of Automation and Computer Science*
*Technical University of Cluj-Napoca*
Cluj-Napoca, Romania
claudia.pop@cs.utcluj.ro

Razvan Zanc
*Global Business Services*
*IBM Romania*
Cluj-Napoca, Romania
razvan.zanc@ibm.com

Ionut Anghel
*Computer Science Department*
*Faculty of Automation and Computer Science*
*Technical University of Cluj-Napoca*
Cluj-Napoca, Romania
ionut.anghel@cs.utcluj.ro

Marcel Antal
*Computer Science Department*
*Faculty of Automation and Computer Science*
*Technical University of Cluj-Napoca*
Cluj-Napoca, Romania
marcel.antal@cs.utcluj.ro

Ioan Salomie
*Computer Science Department*
*Faculty of Automation and Computer Science*
*Technical University of Cluj-Napoca*
Cluj-Napoca, Romania
ioan.salomie@cs.utcluj.ro

*Abstract* - Decentralized management and coordination of energy systems are emerging trends facilitated by the uptake of the Internet of Things and Blockchain offering new opportunities for more secure, resilient, and efficient energy distribution. Even though the use of distributed ledger technology in the energy domain is promising, the development of decentralized smart grid management solutions is in the early stages. In this paper, we define a layered architecture of a blockchain-based smart grid management platform featuring energy data metering and tamper-proof registration, business enforcement via smart contracts, and Oracle-based integration of high computational services supporting the implementation of future grid management scenarios. Three such scenarios are discussed from the perspective of their implementation using the proposed blockchain platform and associated challenges: peer to peer energy trading, decentralized management, and aggregation of energy flexibility and operation of community oriented Virtual Power Plants.

## I. Introduction

Decentralized management and coordination of energy systems are emerging trends in the energy landscape which facilitates collective actions to improve the grid resilience and assure the self-supply of the local energy demand [1]. At the same time, it allows considering the local communities aggregated as Virtual Power Plants (VPPs), as the key, yet missing stakeholder, able to identify energy needs, take own proper initiatives and bring people together to achieve common goals with a view to increase decarbonization of local territories [2].

In support to this transition, utility companies have defined demand response (DR) programs providing the possibility for prosumers (producers and consumers) to play a significant role in the operation of the electricity grid by shaping and coordinating their energy generation and / or flexible demand to deliver energy or ancillary services and obtaining in exchange financial benefits [3]. However, due to the proliferation of more and more dispersed and small-scale distributed energy resources (DERs), more significant energy flexibility may be offered and delivered to system operators, if such flexibility is aggregated at scale. Virtual aggregation models have emerged aiming to enlist and coordinate small scale prosumers to trade their combined flexibility to utilities taking a percentage of service incentives as compensation and at the same time to sell their surplus energy generation in wholesale energy markets [4], [5].

Lately, in the Information and Communications Technology (ICT) domain, the research and industry have gained a lot of interest in the blockchain technology and its potential in decentralizing the management of complex energy systems [6],[7], [8]. Blockchain-based platforms have recently emerged in a range of sectors as effective ways for reducing costs, improving control, and allowing small size suppliers to compete with large traditional ones. Blockchain-based infrastructures may be conveniently used for those domains characterized by high demand variability and diversity and low production economies of scale, which also occur in power networks and energy markets being aligned with the rise and deployment of small-scale renewable sources. The advantage brought by blockchain is the tamper-proof log of all energy transaction while the self-enforcing smart contracts can implement the rules and constraints that need to be verified and agreed upon by all peer nodes from the energy network [9]. These contracts are registered in ledger's blocks and triggered by transaction calls that require each node to update its state based on the results obtained after running the smart contract. Even though the blockchain and smart contracts are considered the emerging technology that can be used for a decentralized grid topology, few state-of-the-art approaches are discussing their applicability for the management of flexibility or VPPs [10], [11], [12].

The rest of this paper is structured as follows. Section II describes the blockchain technological enablers organized in an architecture of smart grid decentralized management, Section III presents the smart grid advance management scenarios and their implementation, while Section IV presents conclusions and future work.

## II. BLOCKCHAIN BASED SYSTEM FOR GRID MANAGEMENT

Figure 1 presents an overview of the main blockchain technological enablers that support the implementation of future smart energy grid management scenarios.

At the **Prosumer Layer**, Internet of Things (IoT) smart energy meters devices are deployed for dealing with the energy data monitoring, acquisition and registration in the blockchain.

The **Protocol and Network Layer** contains the main elements of the distributed ledger technology. The Protocol Layer features solutions for energy registration as asset, transactions, data structure and privacy. The Network Layer features solutions for creating the peers' network, blocks replication and consensus-based ledger state validation [9]. The latter ensures the tamper-proof character of the data stored within the network through a Proof-of-Authority consensus mechanism. Whenever a new prosumer joins the blockchain network a new account is created by generating a pair of public-private keys. The public key is used to create the address that represents the prosumer in the blockchain based management system.

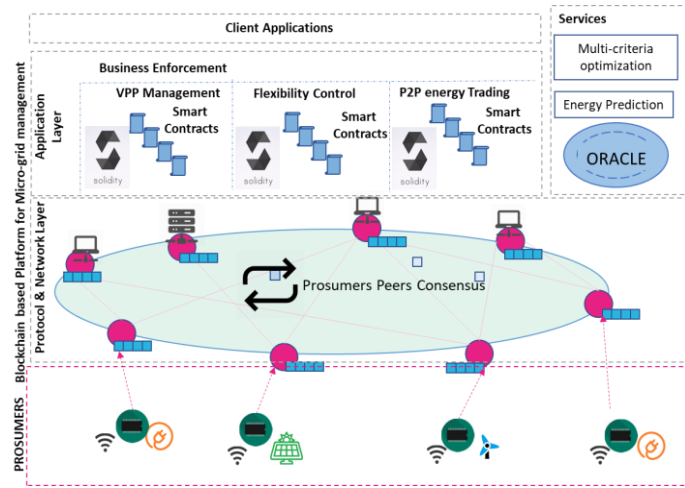

Fig. 1. Design of a blockchain based smart grid management system

A network-connected device is installed on-premises, which will act on behalf of the client, signing energy transactions and forwarding monitored energy data. For each prosumer IoT energy metering devices an associated IoT smart contract is created on-chain through a deploy transaction. Each smart contract will store immutable metadata about the IoT monitoring device (device type, measurement type, etc.) and mutable information regarding the device's monitored state. When the new energy monitored data is available to be registered on-chain, a transaction will be signed having as a payload the monitored data and as a receiver the address of the associated smart contract. The energy transaction is issued on-chain to be mined in the future blocks also providing the associated signature for future authentication and validation of the transaction. The monitored data provided in each transaction will update the state of the corresponding smart contract. This transaction and the associated state update will be stored in a block, benefiting from all the advantages brought by the technology in terms of availability, reliability, consensus, and immutability.

The **Application layer** assures the implementation and enforcement of business logic rules associated with different smart grid management options at the level of smart contracts deployed on-chain. This opens new opportunities, since the validators of the business rules are the actual prosumers and the smart grid management processes do not have to rely on trusted third parties anymore.

The **Computational Scaling** is assured via **Oracle** a standalone component waiting for computation triggering events. It is custom implemented so that depending on the request can connect to different services. The energy prediction service may offer fine grained energy forecasting functionalities to prosumers, aggregators or Distribution System Operator (DSO) that are useful for implementing the foreseen smart grid management scenarios. For example the forecasting of energy consumption and production on different time horizons and granularities (e.g. day-ahead: energy values for the next 24 hours with a granularity of 1 hour; or intra-day: energy values for the next 4 hours with granularity of half an hour) are used for defining the energy bids and offers on the peer to peer (P2P) energy market. The prediction algorithms are trained based on the monitored energy traces acquired using smart energy meters. Flexibility forecasting features can be used by the aggregators and their enrolled prosumers to better define and manage the DR programs based of flexibility shifting in time. A multi-criteria optimization service may provide features for optimal matching the energy bids and offers on the P2P energy market and may determine the clearing price, for identifying the optimal subset of prosumers and their flexibility energy profiles to deliver a specific aggregated energy flexibility to an aggregator and finally for determining the optimal coalitions of prosumers in a VPP.

The services response will be integrated into fully customizable smart contracts allowing the injection of various signals, local constraints and multi value objectives potentially addressing a variety of management scenarios that have the potential to bring an improved, fair share of benefits to prosumers and to the energy system. Also, by leveraging on smart contracts, a decentralized energy and financial settlement process can be implemented assuring tracking in a near real-time fashion the identity of the prosumers, deviations in delivery and getting the billing process closer to the real-time.

## III. GRID MANAGEMENT SCENARIOS

By leveraging on the blockchain-based platform for smart grid management described in the previous section, the following scenarios can be enforced:

- the decentralized flexibility control for distributed provisioning of energy flexibility from DSO to aggregators and further to their enrolled prosumers;
- the P2P energy trading via an energy marketplace;
- the creation of coalitions of prosumers, that together can match the requests of the grid by delivering cooperative loads as VPPs.

### A. Flexibility Aggregation and DR

Resource aggregators in the electricity markets are acting like third-party intermediates between the energy stakeholders, and prosumers of electricity from a micro-grid. Energy aggregators leverage on final customers by managing their electrical loads throughout peak periods or changing certain actions to other periods of the day when the energy price is lower [13]. Thus, the prosumers will offer and trade their flexibility in terms of loads modulation.

In such a management scenario, the flexibility aggregators must implement methods that enable their communication with the prosumers as effectively as possible. This is well featured in our proposed blockchain based architecture, the energy transactions being stored in blocks that are replicated while the smart contract will self-enforce each time the state of the distributed ledger is changed.

The deployment and integration of IoT metering devices at prosumer sites and with the blockchain smart contracts will enable efficient control of electrical loads and provide powerful communication features. This will support two-way communication between the smart grid players such as customers, retailers, aggregators and system operator facilitating the implementation of communication models such as OpenADR [14].

The self-enforcing smart contracts will manage the levels of energy demand flexibility, associating incentive or penalties rates. The Oracle based integration of services will allow the determination of DR signals to be injected in smart contracts to control the power load and provide power usage rates of different devices, while also reporting important environmental conditions.

The energy aggregators conduct a cost / benefit analysis for each of their prosumers in order to be certain that their contributions to the DR programs are positive in a long run. The reward of the prosumers depends on the amount of flexibility they have shifted in the program. This is assessed in relation to their Baseline Profile, which is the standard energy demand of a client outside of a DR program. The baseline profile is crucial for evaluating the power variation due to DR signals. If the baseline is not clearly defined uncertainty might arise to prosumers decreasing their enrolment with the programs. In the blockchain approach the smart contracts define the prosumers baseline profile and expected adjustments in terms of the amount of energy flexibility to be shifted. Smart energy meters and associated smart contracts allow the analysis of energy variation due to various signals received from the aggregator, which is a must for the successful billing and reward. Moreover, the blockchain tamper registration of energy data will allow the auditing of both energy and financial settlement process by all the peers in the network, thus increasing the trust.

The aggregators provide for the DSOs improved resilience and lower management costs by including multiple prosumers to satisfy the limits of grid equipment through peak demand intervals. The power producer can offer incentives to consumers through an aggregator to participate in a DR program during high renewable energy generation periods by increasing their consumption loads or during peak demand periods by decreasing their consumption loads. The aggregators can also improve the quality of the voltage in the grid by investing in compensators and building a properly coordinated and decentralized controller using the existing devices such as on-load tap changers and integrating better tools/services for capacity forecasting [15]. In our blockchain solution, congestion points determined by the DSO are communicated using smart contract to the aggregators which are activated and will contract prosumers connected to the congestion point to offer flexibility. Using the flexibility order aggregators will leverage on prosumer level smart contracts to adjust the load of their prosumers as to fulfil the flexibility needs.

In Europe, aggregators can offer to network operators a comprehensive resilient portfolio of defined and dispatchable power, by joining into household consumers. The versatility can be used to smoothen the intermittent nature of renewable energy, which varies on an hourly basis. The regions of the grid that have a higher penetration level of wind and sun require more levels of generation [16].

### B. P2P Energy Trading

P2P electricity trading allows prosumers in a micro-grid to sell their extra electricity generated to the consumers aiming to maximize the uptake of renewable energy. This is technological challenging because the assumption that multiple prosumers will trade electrical power among them without the governance of a central authority makes the process subject to potential fraud. In our blockchain based smart grid management platform this will be mitigated by leveraging on the distributed ledger technology (DLT) feature of storing the energy transaction in a tamper proof, replicated and trackable manner.

Another major challenge that rises in a power-trading platform that is heavily used, is that the decision-making method for the parameters can collide with the concerns of different prosumers. The process of exchanging electrical power is distinct from other exchange processes because the prosumers are all part of a network, which has strong technical restrictions on power transfer. The scenario where multiple stakeholders request services from a prosumer, with different objectives in mind is also a challenge that is usually encountered and must be carefully managed using various pricing strategies to prioritize stakeholder requests to not overload the power network or encountering power losses

[17]. This is addressed in our blockchain based architecture by using external services integrated via Oracles for applying complex logic or activity evaluation upon the registered energy values and custom logic through the execution of smart contracts on the fetched values.

The defined architecture assures the implementation of a distributed network in which the prosumers of the micro-grid share resources (i.e. energy of flexible assets) with each other. The shared resources can be reached by any network members without the mediation of any third-party entities.

The physical energy network assists the transferal process of electrical power from sellers to buyers after the financial transaction is finished in the virtual layers. This physical network can be implemented using a traditional distributed-grid or using an extra microgrid in combination with the traditional distributed-grid. The physical layer must provide the devices needed for the information exchange among prosumers and with the blockchain based layer. The financial transaction among different prosumers in the virtual platform does not guarantee the physical distribution of electricity. The transaction is in fact a signal sent from the buyers to their prosumers to start the injecting energy into the grid [18].

A key element of peer to peer energy trading is the blockchain based information system. It connects market members to interact with each other for engaging in power trading and offers a market platform to network members. Also monitors the evolution of the electricity market, and also set limitations to network members to guarantee network security and safety [19].

Pricing mechanisms are introduced within market operations via Oracles based integration of external services and are used for balancing the power supply and demand. Pricing strategies used in P2P trading are different from other markets because the price needs to consider the state of the energy inside the P2P network (a higher volume of power should lower the rate and opposite). For matching the energy demand and production it is crucial to set the technical constraints of the micro-grid and to consider the prediction of energy generation and consumption at future time windows.

For P2P trading to work as expected, enough prosumers need to be enrolled with the blockchain system and some of those members need to be able to produce energy to meet the demand. Due to its transparency and decentralization property, the DLT has the potential of increasing the level of prosumers involvement in energy trading removing some of the market entry barriers. To become operational, the P2P energy trading must be regulated. In our approach the rules are implemented in the market design via the customizable smart contracts that implement the business rules in the Application layer.

Finally, in a decentralized P2P market, prosumers should be able to negotiate with each other in order to decide the trading details without a central administration [20]. The business enforcement via smart contracts consider the agreements between prosumers, as well as the energy stream balance and market of forecasting process uncertainty.

*C. Community oriented VPP Management*

This distributed management scenario addresses the increasing need to optimize the output from multiple local generation assets (i.e. wind-turbines, small hydro, photovoltaic, back-up generators, etc.) that primarily serve local communities, but also have export connections to power distribution network [21], [22]. The goal is to provide the technological means for considering in near real time fashion the distributed generation of electricity of multiple types with a view of creating optimal coalitions of prosumers to provide a reliable aggregated power supply. The concept of a VPP can be reduced to a cluster of scattered prosumers, controllable flexible assets, and energy storages systems, all aggregated into one virtual unit that has the main objective to cooperate to contribute to pre-defined smart grid sustainability objectives. The blockchain based management system that has the main responsibility of registering, tracing and coordinating the energy volumes that come from the peer members of the VPP. Being built on top of the P2P local energy trading and decentralized flexibility management, this scenario will allow a set of prosumers to be aggregated and ultimately participate on national energy markets [23] such as a flexibility provider on the ancillary services market or capacity provider on the wholesale market (see Figure 2). The VPP will be able to maximise utilisation and revenues from Renewable Energy Sources (RES) by leveraging on the self-enforcing smart contracts to adapting its capacity to the optimum paying service thus maximizing the associated profit.

In this context, the use of self-enforcing smart contracts for the aggregation of generation, storage and flexible assets assures not only the consideration of assets specific constraints but also the community level objectives and global objective functions associated with a service to be provided. This is due to the heterogeneous nature of the smart grid energy assets having different energy profiles and different response times to control signals thus allowing their operation to be optimal distributed and tracked by using the smart contracts. At the same time, leveraging on their amount of energy generation by grouping the distributed energy sources in a single portfolio may generate benefits.

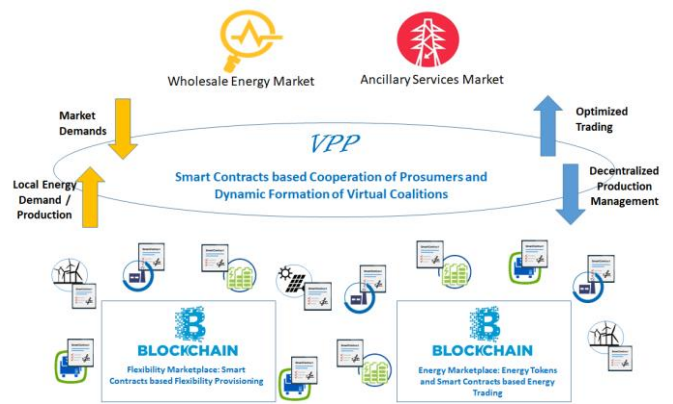

Fig. 2. Virtual Power Plants management using blockchain

In this case would be a VPP that is active on the wholesale market by operating on a profit maximizing function while supporting the need of local grid prosumers and consumers. In a similar manner the VPP could leverage on the aggregated flexibility profiling of its individual peers to provide flexibility services for a DSO. The smart contracts are used for tacking and regulating the baseline of individual prosumers to estimate the flexibility potential generation of storage assets. This will allow for automatic and decentralized assignment of new sites / assets to an outbound service based on the asset's particularities such as response time, sync times and maximum dispatch period (i.e. offering power factor regulation [24]).

In this context prosumers and assets segmentation in flexibility bands via Oracle based integration with external services on the generation side is crucial for allowing different peers to be used simultaneously for various services. The VPP operation is following the services associated multi-objectives by using energy records stored and blockchain transactions and finalized by near real-time validation, settlement and remuneration. Part of these activities are enabled at the energy resource level by advanced smart meters enacting real-time observability at high accuracy of voltage and power measurements, energy resource control through the self-enforcing smart contracts local intelligence and minute-based load profiles of energies digital fingerprinted and stored in a tamper-evident manner in the distributed ledger.

## IV. CONCLUSIONS

In this paper we propose the architecture of a blockchain based platform for decentralized management of smart energy grid. The platform features the integration of IoT energy metering devices and tamper proof registration of monitored data, business rules enforcement using smart contracts and Oracles for integration of high computation services. The implementation and challenges of three management scenarios is discussed: (1) the peer to peer energy trading, (2) decentralized and aggregation of flexibility and finally (3) the operation of a VPP a decentralized coalition of prosumers and energy assets. The proposed architecture and management scenarios may successfully address key challenges brought by the uptake and deployment of small-scale renewable energy sources contributing to the grid decarbonization.

ACKNOWLEDGMENT

This work has been conducted within the BRIGHT project Grant number 957816, co-funded by the European Commission as part of the H2020 Framework Programme (H2020-LC-SC3-2018-2019-2020).